\documentclass[11pt]{article}

\usepackage{fullpage}
\usepackage{subfig}
\usepackage[usenames,dvipsnames]{xcolor}
\usepackage[colorlinks,citecolor=blue,linkcolor=BrickRed]{hyperref}
\usepackage{times}
	\usepackage{latexsym,graphicx,epsfig,color}
\usepackage{amsfonts,amssymb,amsmath,amsthm,amstext}
\usepackage{thm-restate}
\usepackage{enumitem}
\usepackage{url,setspace}
\usepackage{multirow}
\usepackage{rotating}
\usepackage{makeidx}
\usepackage{tikz}
\usepackage{enumitem}
\usepackage{xspace}
\usepackage{algorithm,algpseudocode}
\usepackage{bm}
\usepackage{changepage} 	
\usepackage{cleveref}
\usepackage{mathabx}  
\usepackage{accents}

\newcommand{\IGNORE}[1]{}
\newcommand{\mc}[1]{\mathcal{#1}}
\newcommand{\defeq}{\stackrel{\mathrm{def}}{=}}
\newcommand{\Opt}{\mathrm{OPT}}

\DeclareMathOperator{\val}{val}
\DeclareMathOperator{\adap}{adap}
\DeclareMathOperator{\greedy}{greedy}

\DeclareMathOperator{\alg}{alg}

\usetikzlibrary{decorations.markings}
\usetikzlibrary{arrows}
\tikzstyle{block}=[draw opacity=0.7,line width=1.4cm]
\tikzstyle{graphnode}=[circle, draw, fill=black!20, inner sep=0pt, minimum width=6pt]
\tikzstyle{point}=[circle, draw, fill=black!30, inner sep=0pt, minimum width=1pt]
\tikzstyle{input}=[rectangle, draw, fill=black!75,inner sep=3pt, inner ysep=3pt, minimum width=4pt]
\tikzstyle{unmatched}=[graphnode,fill=black!0]
\tikzstyle{shaded}=[graphnode,fill=black!20]
\tikzstyle{matched}=[graphnode,fill=black!100]  	
\tikzstyle{matching} = [ultra thick]
\tikzset{
    >=stealth',
    pil/.style={
           ->,
           thick,
           shorten <=2pt,
           shorten >=2pt,}
}
\tikzset{->-/.style={decoration={
  markings,
  mark=at position .5 with {\arrow{>}}},postaction={decorate}}}

\makeindex

\DeclareMathOperator{\argmax}{arg max}

\newtheorem{theorem}{Theorem}[section]
\newtheorem{claim}[theorem]{Claim}

\newtheorem{lemma}[theorem]{Lemma}

\theoremstyle{definition}

\newtheorem{defn}[theorem]{Definition}

\usepackage{thmtools,thm-restate} 

\newcommand{\E}{\mathbb{E}}

\def \reals {\mathbb{R}}

\def \integers {\mathbb{Z}}

\newcommand{\SMP}{\ensuremath{{\sf SMP}}\xspace}

\newcommand{\no}{\texttt{no}\xspace}
\newcommand{\yes}{\texttt{yes}\xspace}

\def\adap{\ensuremath{{\sf adap}}\xspace}

\def\alg{\ensuremath{{\sf alg}}\xspace}

\newcommand{\initOneLiners}{
    \setlength{\itemsep}{0pt}
    \setlength{\parsep }{0pt}
    \setlength{\topsep }{0pt}}

\newcounter{note}[section]

\newcommand{\myvec}{\mathbf}
\newcommand{\x}{\myvec{x}}
\newcommand{\vX}{\myvec{X}}
\newcommand{\T}{\mathcal{T}}
\newcommand{\calF}{\mathcal{F}}
\newcommand{\calP}{\mathcal{P}}

\newcommand{\univ}{V}

\allowdisplaybreaks

\title{(Near) Optimal Adaptivity Gaps for \\
 Stochastic Multi-Value Probing}

	\author{Domagoj Bradac\thanks{
	(domagoj.bradac@gmail.com)
  Department of Mathematics, Faculty of Science, University of Zagreb.
	}
	\and Sahil Singla\thanks{
        (singla@cs.princeton.edu)
        Department of Computer Science,
        Princeton University.
        Most of this work was done when the author was a graduate student at Carnegie Mellon University.
        }
	\and Goran Zuzic\thanks{
        (gzuzic@cs.cmu.edu)
        Computer Science Department,
        Carnegie Mellon University.
        }
}
  
\date{ \today}

\begin{document}
\maketitle

\setlength{\abovedisplayskip}{3pt}
\setlength{\belowdisplayskip}{3pt}

\begin{abstract}{
    Consider a kidney-exchange  application where we want to find a max-matching in a random graph. To find whether an edge $e$ exists, we need to perform an expensive test, in which case the edge $e$ appears independently with a \emph{known} probability $p_e$. Given a budget on the total cost of the tests, our goal is to find a testing strategy that maximizes the expected maximum matching size.

    The above application is an example of the stochastic probing problem. In general the optimal stochastic probing strategy is difficult to find because it is \emph{adaptive}---decides on the next edge to probe based on the outcomes of the  probed edges. An alternate approach is to show the \emph{adaptivity gap} is small, i.e., the best \emph{non-adaptive} strategy  always has  a   value close to the best adaptive strategy. This allows us  to focus on designing  non-adaptive strategies that are much simpler. Previous works, however, have  focused on Bernoulli random variables that can only capture  whether an edge appears or not. In this work we introduce a multi-value stochastic probing problem, which  can  also model situations where the weight of an edge  has a  probability distribution over multiple values.

    Our main technical contribution is to obtain (near) optimal bounds for the (worst-case) adaptivity gaps for multi-value stochastic probing over prefix-closed constraints. For a monotone submodular function, we show the adaptivity gap is at most $2$ and provide a matching lower bound. For a weighted rank function of a $k$-extendible system (a generalization of  intersection of $k$ matroids), we show the adaptivity gap is between $O(k\log k)$ and $k$. None of these results were known even in the Bernoulli case where both our upper and lower bounds also apply, thereby resolving an open question of Gupta et al.~\cite{GNS-SODA17}.
}\end{abstract}

\thispagestyle{empty}
\newpage

\setcounter{page}{1}


\section{Introduction} \label{sec:intro}

Consider a kidney-exchange application where we want to find a maximum matching in a random graph.
To find whether an edge $e$ exists, we need to perform an expensive test, in which case the edge $e$ appears independently with a known probability $p_e$. Given a budget on the total cost of the tests, our goal is to design a testing strategy that maximizes the expected size of the found matching.


The above application can be modeled as a constrained \emph{stochastic probing}  problem~\cite{ANS-WINE08,GN-IPCO13,ASW14,GNS-SODA16,GNS-SODA17}. In this setting, we are given a universe $\univ$ of \emph{elements} (e.g., the set of all possible edges), each with an \emph{activation} probability $p_v$ for $v\in V$ (e.g., the probability an edge exists). We define a random set $A \subseteq \univ$ of \emph{active} elements that contains every $v$ independently with probability $p_v$. A \emph{probe} at $v$ reveals whether $v \in A$ or $v \not \in A$, and we are only allowed to probe certain \emph{feasible subsets} $S \in \calF \subseteq 2^\univ$ (e.g., subsets of edges whose tests fit in our budget). Our goal is to design a \emph{probing strategy} to find a feasible set $S \in \calF$ of elements to maximize $\E_A[ f(A \cap S) ]$, where $f$ is some combinatorial function $f: 2^\univ \rightarrow \reals_{\geq 0} $ (e.g., the cardinality of the maximum matching). 
Notice our probing strategy could be \emph{adaptive}, i.e., we could decide which element to probe next based on the outcomes of already probed elements.

Besides matching~\cite{CIKMR-ICALP09,BGLMNR-Algorithmica12}, stochastic probing has applications for stochastic variants of several other combinatorial problems. E.g., it can be used  for  Bayesian mechanism design problems~\cite{GN-IPCO13},  robot path-planning problems~\cite{GNS-SODA16,GNS-SODA17}, and stochastic set cover problems that arise in database applications~\cite{LiuPRY-SIGMOD08,DHK-SODA14}.
As observed in these prior works, the optimal strategy for stochastic probing can be represented as a binary \emph{decision tree} where each node represents an element of $\univ$: You first probe the root node element, and then depending on whether it is active or inactive, you either move  to the right or the left subtree. In general, such an optimal decision tree can be exponentially sized and is hard to describe. We do not even understand how to capture it for very simple functions and constraints (e.g., the $\max$ function with cardinality constraints~\cite{FLX-ICALP18}).

\begin{figure} 
\begin{center}
\includegraphics[width=0.55 \textwidth]{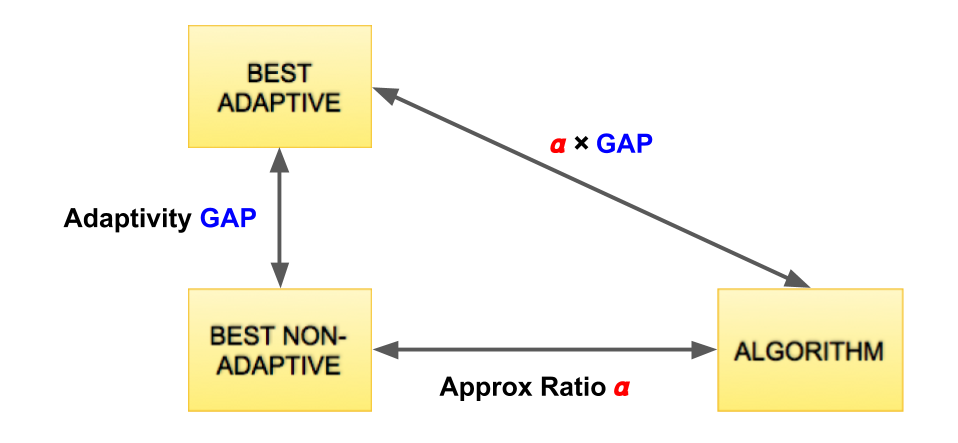}
\end{center}
\vspace{-0.5cm}
\caption{An $\alpha$-approximation to the best non-adaptive solution implies an  $(\alpha \cdot \mathrm{GAP})$-approximation to the best adaptive algorithm, where $\mathrm{GAP}$ is the adaptivity gap.} 
\label{fig:AdapGaps}
\end{figure}

An alternate approach is to focus on \emph{non-adaptive} strategies. Such a strategy commits to probing a feasible set $S\in \calF$ in the beginning, irrespective of which of these elements turn out active. A non-adaptive strategy has several benefits: (a)~it is easy to represent since we can just  store the set $S$, (b)~it is easy to find for many classes of functions and constraints (e.g., submodular functions over intersection of matroids~\cite{CVZ-SICOMP14}), and (c)~it is parallelizable because we do not need feedback. 
The concern is that the expected value of the optimal non-adaptive strategy might be much smaller than that of the optimal adaptive strategy. This raises the (worst-case instance) \emph{adaptivity gap} question: What is the maximum ratio between the expected values of the optimal adaptive and the optimal non-adaptive strategies for stochastic probing? If this ratio is small then we can focus on non-adaptive strategies and reap its benefits with only a small loss in value (see Figure~\ref{fig:AdapGaps}). 

Since  for general combinatorial functions or constraints the adaptivity gaps can be made arbitrarily large, we need to consider  special classes of functions and constraints.
In a surprising result, Gupta et al.  prove that  for any \emph{monotone submodular} function  and any \emph{prefix-closed constraints}\footnote{Prefix-closed constraints stipulate that any prefix  of a feasible probing sequence is also feasible.  This class contains any downward-closed/packing constraint.}, the adaptivity gap is at most $3$~\cite{GNS-SODA17}. 
 The best known lower bound in this setting, however, is only $\frac{e}{e-1}\approx 1.58$ due to Asadpour et al.~\cite{ANS-WINE08}.
This leaves open the following  question:
\vspace{-0.3cm}
\begin{quote}
\emph{For stochastic probing, what is the (worst-case) adaptivity gap for monotone submodular functions over prefix-closed constraints?}
\end{quote}
\vspace{-0.3cm}
We show that both the previously known upper bound of $3$ and the lower bound of $\frac{e}{e-1}$ are not tight. Instead,  the adaptivity gap is exactly $2$.

One might notice that submodular functions do not capture the max-matching function used to model kidney-exchanges. This motivates us to consider more general combinatorial functions; in particular, we study the weighted rank function of a $k$-extendible system (defined in \S\ref{sec:prelims}). This class generalizes intersection of $k$-matroids~\cite{mestre2006greedy}, e.g., a $2$-extendible system captures matching in general graphs (unlike intersections of two matroids). Our goal is to bound the adaptivity gap for such functions over arbitrary prefix-closed constraints.

A major drawback of the stochastic probing model is that it only considers Bernoulli random variables. One would ideally allow for more modeling power by permitting the outcome of a probe to be  a non-binary value. For example, in the kidney-exchange application, one might desire to summarize an edge probe by the risk involved in performing the match: a value of $0$ describes an impossible match, a value of $1$ indicates a safe match, and the possibilities in between are represented by intermediate values.
Notice that the optimal adaptive strategy is still a decision tree; however, it may no longer be binary.

The main contributions of this paper are (1) a model that extends the binary stochastic probing to the multi-value setting, (2) the exact calculation of the adaptivity gap for stochastic probing of monotone submodular functions (in both the binary and multi-value setting), and (3) a nearly-tight adaptivity gap for stochastic probing of weighted rank functions over $k$-extendible systems.

\subsection{Overview of Results}

Our conceptual contribution is to present a generalization of the stochastic probing model to \emph{stochastic multi-value probing} (\SMP) described in \S\ref{sec:prelims}. Roughly, the idea is that each element has $t$ potential types, and a probe reveals which one of its types it takes. This trivially captures stochastic probing for $t=2$, where the two types are active and inactive. In general these different types can be used to model different weights of an element, or to even encode different kinds of complementary relationships in the element values. 

Although the \SMP model is more general than the stochastic probing model, our main  technical result in \S\ref{sec:submod} is that for monotone submodular functions the adaptivity gap is  bounded by $2$. We also give a matching lower bound which proves this cannot be further reduced. This is despite the fact that the optimal decision tree for \SMP may no longer be binary. 

\begin{restatable}{theorem}{submodMain}\label{thm:SubmodAdapGap}
The adaptivity gap for \SMP  where the constraints are prefix-closed and the function is   monotone non-negative submodular  is exactly $2$.
\end{restatable}

Since \SMP is strictly more general than stochastic probing,  Theorem~\ref{thm:SubmodAdapGap} also improves the previously  known upper bound of $3$ for monotone submodular stochastic probing. In fact, our lower bound  \SMP instance in Theorem~\ref{thm:SubmodAdapGap} is Bernoulli. Thus it  resolves an open question of~\cite{GNS-SODA17} of finding the optimal adaptivity gaps for submodular stochastic probing.

Our main  technical result in \S\ref{sec:kSystem} is that the adaptivity gap for weighted rank function of a $k$-extendible system is $\tilde{\Theta}(k)$.

\begin{restatable}{theorem}{kSystemMain}\label{thm:kSystemAdapGap}
The adaptivity gap for \SMP  where the constraints are prefix-closed and the function is  a weighted rank function of a $k$-extendible system is between $k$ and $O(k \log k)$. Moreover, for unweighted rank functions, the adaptivity gap is between $k$ and $2k$.
\end{restatable}

Since the weighted rank of function of intersection of $k$-matroids is a $k$-extendible system, Theorem~\ref{thm:kSystemAdapGap} implies as a corollary that the adaptivity gaps for this class is at most $\tilde{\Theta}(k)$. This improves the previously best known upper bound for intersection of $k$ matroids of $O(k^4\cdot \log n)$ due to Gupta et al.~\cite{GNS-SODA16}. 
We also give an  $\Omega(\sqrt{k})$-lower bound in this setting.

\subsection{Techniques and Challenges}\label{sec:techniques}

In this section we outline our main techniques and challenges for \SMP adaptivity gaps.
 
\noindent \textbf{Submodular Functions:} To prove a small adaptivity gap, we need to show existence of a ``good'' non-adaptive solution. A priori it is not clear how to construct such a solution, e.g., LP based approaches do not extend beyond matroid constraints because of large integrality gaps. 
Since we only need to show \emph{existence}, we can  assume  the optimal (exponential sized) decision tree is known.
A crucial idea of~\cite{GNS-SODA16} is  to perform a random walk on this optimal  decision tree (with probabilities given by the tree) and probing elements on the sampled root-leaf path. 
 In other words, consider a  non-adaptive strategy that randomly chooses a root-leaf path in the decision tree with the same probability as the optimal adaptive strategy.
 While this idea is natural in hindsight, its analysis for the non-adaptive strategy has been challenging.

In~\cite{GNS-SODA16}, the authors use  Freedman's inequality---linear functions are ``well-concentrated'' for a martingale---to  argue that \emph{simple} submodular functions are  well-concentrated. This step requires  massive union bounds over a polynomial number of linear functions, which loses logarithmic factors. To overcome this super-constant loss, in~\cite{GNS-SODA17} the authors use an inductive approach and induct over subtrees where in each step a  \emph{stem}---the all-{\large no} path---is observed. A ``stem lemma'' allows them to argue that for every stem the expected value of the non-adaptive algorithm is within a factor $2$ to the expected adaptive strategy. Finally,  they ``stitch'' back the stem for induction by using submodularity, overall losing a factor of $3$.

In this work, to prove the improved adaptivity gap of $2$ in \Cref{thm:SubmodAdapGap}, our  insight is to modify the above induction to observe a \emph{single node} at each step (instead of a \emph{stem} as in~\cite{GNS-SODA17}). While we still induct over subtrees, this allows us to avoid any additional loss due to the stitching step. This induction turns out to be nontrivial because the adaptive and  non-adaptive strategies can observe different types of the root element. In other words,  although the non-adaptive random walk strategy follows the distribution of root-leaf paths of the adaptive strategy, it has to independently re-sample (re-probe) all the nodes on the chosen path.
This hinders a direct application of induction as the marginal values in the subtrees change between the two strategies. We remedy this issue using two main ideas. First, we compare the non-adaptive strategy to a ``super-strategy'' that can choose from both the elements chosen by the adaptive and the non-adaptive strategies. (This is also the intuition for the gap of $2$ since the ``super-strategy'' has two chances to sample an element.) 
 Second, the non-adaptive strategy forfeits any potential future value that the adaptive strategy gained at the root but the non-adaptive missed due to re-sampling. (This can be done by contracting the element sampled by the adaptive strategy without receiving its value.) Notice that both these steps are pessimistic and hence give a valid upper bound on the adaptivity gap.  
Together these ideas suffice to match the marginal values in the subtrees and apply induction without the stiching step, yielding an adaptivity gap of $2$. Our lower bounds in \S\ref{sub_lower_bound} show examples where the super-strategy does not have any advantage over the adaptive strategy. Thus the adaptivity gap of $2$ is optimal.

\noindent \textbf{Rank Functions:} A technical challenge in extending the above \emph{inductive} approach to   $k$-extendible system rank functions is that their marginal values do not belong to the same class. Namely, after contracting an element, the marginal value of a submodular function is  submodular but the marginal value of a  $k$-extendible system rank function may not even be subadditive.  
To overcome this, we  first focus on \emph{unweighted} rank functions. Instead of directly comparing the non-adaptive strategy to the adaptive strategy, our insight is to compare it to a \emph{greedy procedure}. We show that this  greedy procedure is a $k$-approximation to the adaptive strategy. Moreover, we show it has  a notion of a marginal value. This allows us to compare the non-adaptive strategy to the greedy procedure in a similar way as for submodular functions, by losing another factor of $2$. 
Our lower bound in \S\ref{sec:lowerKSystem} shows that the factor $k$ loss in comparing to a greedy procedure is unavoidable, thereby making our analysis tight up to constants.


Finally, the challenge in proving \Cref{thm:kSystemAdapGap} for \emph{weighted} $k$-extendible system rank functions  is that  the greedy procedure only guarantees  a $k$-approximation if we go in the order of decreasing weights. Instead,  our adaptivity gap proofs  only work when we are greedy in the root-to-leaf path order. One  way around this is to partition the elements into $O(\log n)$ exponentially weighted \emph{classes} (e.g., $1,2,2^2, \ldots$) and apply the unweighted argument to the most valuable class. 
Unfortunately, this loses an $\Omega(\log n)$ factor. 
To obtain bounds independent of the universe size $n$, 
our insight is that picking an element in a  class ``removes'' at most $k$ elements from a lower weight class. We can therefore improve the $\log n$ factor loss to a $\log k$ by increasing the gap between successive classes to  $\Omega(k)$. To achieve this we further combine $O(\log k)$ consecutive classes into a ``super-class" (bucket). It is an interesting open question to find if this $\log k$ loss is essential in going from unweighted to weighted $k$-extendible system  rank functions.

\subsection{Further Related Work}
The adaptivity gap of stochastic packing problems has seen much
interest; see, e.g., for knapsack~\cite{DGV-FOCS04,BGK-SODA11,Ma-SODA14}, packing integer
programs~\cite{DGV-SODA05,CIKMR-ICALP09,BGLMNR-Algorithmica12}, budgeted multi-armed
bandits~\cite{GM-STOC07,GKMR-FOCS11,LiYuan-STOC13,Ma-SODA14}, and
orienteering~\cite{GM-ICALP09,GKNR-SODA12,BN-IPCO14}. All except the orienteering
results rely on having relaxations that capture the constraints of the
problem via linear constraints.
For stochastic monotone submodular functions where the probing
constraints are given by matroids, Asadpour et al.~\cite{AN16} bounded
the adaptivity gap by $\frac{e}{e-1}$; Hellerstein et al.~\cite{HKL15}
bound it by $\frac1\tau$, where $\tau$ is the smallest probability of
some set being materialized. Other relevant papers are~\cite{LiuPRY-SIGMOD08,DHK-SODA14}.

The work of Chen et al.~\cite{CIKMR-ICALP09} (see
also~\cite{Adamczyk-IPL11,BGLMNR-Algorithmica12,BCNSX-APPROX15,AGM-ESA15}) sought to maximize the size of a
matching subject to $b$-matching constraints; this was motivated by
applications to online dating and kidney exchange. See also~\cite{RSU-JET05,AR-AER12} for pointers to other work on kidney exchange
problems. The work of~\cite{GN-IPCO13} abstracted out the general problem of
maximizing a function (in their case, the rank function of the
intersection of matroids or knapsacks) subject to probing constraints
(again, intersection of matroids and knapsacks). This was
improved and generalized by Adamczyk et al.~\cite{ASW14} to submodular
objectives. All these results use LPs  or geometric relaxations, and  do not extend to arbitrary packing constraints due to large integrality gaps of the relaxations.




\section{Stochastic Multi-Value Probing  Model 
}\label{sec:prelims}


In this section we formally define our \emph{stochastic multi-value probing} (\SMP) model using the idea of  combinatorial valuation over independent elements. We also discuss some  preliminaries.
 
\subsection{Combinatorial Valuation over Independent Elements}
The  multi-value paradigm is based on the notion of \emph{type}, which  represents different ``values'' an element can take. This leads to combinatorial valuations over independent elements where each  element \emph{independently} takes its type.
Similar notions have been defined before; e.g., see~\cite{RS-SODA17} and  references therein.

\begin{defn}[Combinatorial valuation $\val_{\vX}$ over independent elements]
Consider a finite universe $\univ$ of {elements} and size $n= |\univ|$. Each element $e \in  \univ$ obtains exactly one  \emph{type}  from a finite set $T_e$ according to a given probability distribution $\mc{D}_e$ over  $T_e$.  These types are assigned independently across different elements, i.e., the random vector of types $\vX \in \bigtimes_{e \in V} T_e$ is drawn from the product distribution $ \prod_{e \in \univ} \mc{D}_e$. 
Given a combinatorial function $f : 2^T \to \reals_{\geq 0}$ for $T \defeq \bigcup_{e \in \univ} T_e$, the  valuation  of a set $S \subseteq \univ$ is
\begin{align*}
\val_{\vX} (S) \defeq  f\big(\{ \vX_e \mid e \in S \}\big) = f( \vX_S ),
\end{align*}
where we define $\vX_S \defeq \big\{ \vX_e \mid e \in S \big\}$ to simplify notation.
\end{defn}


For example, in the Bernoulli case  studied in the stochastic probing literature, each element has two types: active and inactive, the distributions $\mc{D}_e$ are Bernoulli, and the valuation function  $\val_{\vX}(S) = f(\{e \in S \mid e \text{ is active} \})$. Another example is the multi-value max-weight matching problem described in the introduction. Here different types of an element (edge) correspond to its different weights and $\val_{\vX}(S)$ is the max-weight matching in the induced subgraph on $S$.

In this work we always assume the  combinatorial function $f  : 2^T \to \reals_{\geq 0}$ satisfies  $f(\emptyset) = 0$ and  is  \emph{monotone}, i.e., $f(A) \le f(B)$ for all $A \subseteq B$. We also assume it belongs to one of the following classes. 
\begin{itemize}[noitemsep,nolistsep]
\item \emph{subadditive} if $f(A \cup B) \le f(A) + f(B)$ for all $A, B \subseteq T$. 
\item \emph{submodular} if $f(A \cup B) + f(A \cap B) \le f(A) + f(B)$ for all $A, B \subseteq T$. For $ S \subseteq T$, the contraction 
 \begin{align} \label{eq:fnContra}
f_S(A) \defeq f(S \cup A) - f(S)
\end{align}
 of a monotone submodular function is also monotone submodular.
\item \emph{weighted rank function} of a family $\calF \subseteq 2^T$ if $f(A) = \max_{B \in \calF} w(A \cap B)$ where $w : 2^T \to \reals_{\ge 0}$ is a  linear function with non-negative weights. When $w$ is the all ones vector (i.e., $w(A) = |A|$),  we call it  the \emph{unweighted rank function} of  $\calF$.
\end{itemize}

In particular, we work with rank functions of two special families $\calF \in 2^\univ$. Subsets in the family are called \emph{independent} subsets. A family $\calF \ni \emptyset$ forms a
\begin{itemize}[noitemsep,nolistsep]
\item \emph{matroid} if  for every $A, B \in \calF$ with $|A| > |B|$ there exists $x \in A \setminus B$ such that $B \cup \{e\} \in \calF$.
\item \emph{$k$-extendible system} if for every $A \subseteq B \in \calF$ and $e \in T$ where $A \cup \{e\} \in \calF$, we have that there is a set $Z \subseteq B \setminus A$ such that $|Z| \le k$ and $B \setminus Z \cup \{e\} \in \calF$. 
\end{itemize}
This latter family is important because it generalizes the family of intersection of $k$ matroids, e.g., a $2$-extendible systems captures general graph matchings  (see~\cite{CCPV-SICOMP11} for further discussion).

\subsection{Adaptive Strategies and \SMP}\label{sec:combinatorialButIndep}
Roughly, the goal of an \SMP problem is to maximize a combinatorial function over independent elements under some ``feasibility constraints". We   define a \emph{probe}  of an element $e\in \univ$ to be an operation that reveals its random type $X_e \in T_e$.  A \emph{probing sequence}  is an ordered sequence of probes  on some elements.

The \SMP problem only allows a family of  probing sequences $\mc{C}$, which are called \emph{feasible}.  We assume minimal properties from this family. Specifically, it is   \emph{prefix-closed}, i.e., for every sequence in $\mc{C}$, each of its  prefix is also in $\mc{C}$.
This prefix-closed family  is  powerful because it generalizes any \emph{downward-closed}  family $\calF$ (i.e., for all $A \in \calF$ and $B \subseteq A$ we have $B \in \calF$) and can also capture precedence constraints.

We now define an \emph{adaptive strategy} which constitutes  a feasible solution for \SMP. The nodes in this tree correspond to probes of  elements

\begin{defn}[Adaptive strategy $\T$]
It is a rooted decision tree where each non-leaf node is labeled with an element $e \in \univ$ and has $|T_e|$ arcs to child nodes. Each arc is uniquely labeled with a type $t \in T_e$. Whenever we encounter a node labeled $e$, the adaptive strategy probes $e$ and proceeds to  the subtree corresponding to the arc labeled  $X_e \sim  \mc{D}_e$. The strategy terminates on reaching a leaf and receives a value of $\val_{\vX} (S(\vX))$,  where $S(\vX) \subseteq \univ$ is the set of probed elements by strategy $\T$ for type vector $\vX$. The objective  is the expected valuation, which we denote by
\begin{align}\label{eq:defnAdap}
\adap(\T, f) \defeq  \E_{\vX}[ \val_{\vX}(S(\vX)) ] .
\end{align}
\end{defn}
Notice, since  $f$ is monotone,  a strategy never gains value by removing a probed element.
We say a  strategy $\T$ is \emph{feasible} for  $\mc{C}$ if every root-leaf path belongs to $\mc{C}$. We  now formally define an \SMP problem.

\begin{defn}[\SMP problem $(\mc{C}, \val_{\vX})$] Given a prefix-closed family of probing constraints $\mc{C}$ and a combinatorial valuation $\val_{\vX}$ over independent elements, an \SMP problem is to find a feasible  adaptive strategy $\T$  to maximize the expected valuation $\adap(\T, f)$.
\end{defn}

\subsection{Non-Adaptive Strategies and Adaptivity Gaps}

A strategy to solve an  \SMP problem  can benefit from adjusting its probing sequence based on the outcomes of the already probed elements. For instance, in the kidney-exchange example if one finds an edge incident to a vertex $u$,  one may choose not to probe any other edges incident to $u$. On the other hand, a strategy that always decides the next probe independent of the outcomes of the probed elements is  called \emph{non-adaptive}. Our goal is to study the largest ratio between  adaptive and  non-adaptive strategies.

\begin{defn}[Adaptivity gap for $\calP$] Let $\calP$ be a class of SMP problems (e.g.,   monotone submodular functions over  prefix-closed constraints). Define the \emph{adaptivity gap} as the largest (worst-case instance) ratio of the optimal adaptive and optimal non-adaptive strategies for a problem $(\mc{C}, \val_{\vX}) \in \calP$, i.e.,
\begin{align*}
  \sup_{(\mc{C}, \val_{\vX}) \in \calP} \frac{\sup_{\T \text{ is feasible in } P} \adap(\T, f)}{\sup_{S \in \mc{C}} \E_{\vX}[ \val_{\vX} (S) ]} .
\end{align*}
Notice that in the denominator $S$ does not depend on $\vX$. 
\end{defn}
The adaptivity gap for a general combinatorial function $f$ is unbounded~\cite{GNS-SODA16}. In this work we focus on  monotone submodular functions and (weighted) rank functions of a $k$-extendible system.
We bound  adaptivity gaps by  analyzing the following natural random walk non-adaptive strategy.  
\begin{defn}[Random walk non-adaptive strategy] \label{def:randomWalk}
For any given adaptive strategy $\T$, there is a corresponding non-adaptive strategy that (virtually) draws a sample $\vX \sim \prod_{e \in \univ} \mc{D}_e$ from the product distribution and traverses $\T$ along the root-leaf path for $\vX$ (i.e., when at a node labeled  $e$, traverse the unique arc labeled $X_e$). Let $S(\vX)$ be the random set of elements probed by such a root-leaf path. The true (non-virtual) types of elements correspond to the vector of outcomes $\vX' \sim \prod_{e \in \univ} \mc{D}_e$. Here $\vX$ and  $\vX'$ are i.i.d. r.v.s. The random walk non-adaptive strategy probes $S$ according to the above distribution and receives the valuation 
\begin{align}\label{eq:defnAlgNA}
\alg(\T, f) \defeq \E_{\vX, \vX'}[ \val_{\vX'}(S(\vX)) ].
\end{align}
\end{defn}

\section{Adaptivity Gaps for a Monotone Submodular
Function}\label{sec:submod} In this section we prove our first main result, the optimal adaptivity gap for submodular functions. In \S\ref{sec:submodUpper} we prove the upper bound and in~\S\ref{sub_lower_bound} we prove the lower bound of Theorem~\ref{thm:SubmodAdapGap}.
\submodMain*

\subsection{Upper Bound of $2$} \label{sec:submodUpper}
Our non-adaptive strategy samples a random root-leaf path  using the optimal adaptive strategy tree $\T$ (\Cref{def:randomWalk}). In other words, it performs a ``dry-run'' of a random walk along the tree without probing anything. In the end it queries all the elements on this random root-leaf path. We argue that its expected value is at least half of the adaptive strategy. We encourage the reader to follow the proof idea outlined in \S\ref{sec:techniques} since algebra can conceal the main ideas.

\begin{proof}	[Proof of the upper bound in \Cref{thm:SubmodAdapGap}]
We induct over the {depth} of the tree $\T$, i.e., for any monotone submodular function $f$ and tree $\T$ of depth at most $d$, we have
\[	\alg(\T,f) \geq \frac12   \adap(\T,f).
\]
The base case for $d=1$ is trivially true because the tree is a single node. For induction, let $e$ be the root node of the optimal decision tree $\T$. Denote by $I \defeq X_e$ the (random) type of element $e$ when probed by the adaptive strategy (and also the virtual type of the non-adaptive strategy), while $R \defeq X'_e$ be the (random) true type when probed by the non-adaptive strategy. Also, let $\T_I$ denote the subtree the adaptive strategy goes to when the root element is in type $I$ and let $f_I$ be the contraction from Eq.~\eqref{eq:fnContra}. This implies
\begin{align} \label{eq:expansionAdapNA}
 \adap(\T,f) = \E_I [f(I) + \adap(\T_I,f_I)]  \qquad \text{  and }  \qquad \alg(\T,f) = \E_{I,R} [f(R) + \alg(\T_I,f_R)] . 
 \end{align}
Now using submodularity and monotonicity of $f$, we  bound the
adaptive strategy  
\begin{align*} \adap(\T,f) & \leq \E_{I,R} [f(I \cup R) + \adap(\T_I,f_{I\cup R})] \\
 &\leq \E_{I,R} [f(I) + f(R) + \adap(\T_I,f_{I\cup R})],
  \end{align*} 
where the last inequality uses that every monotone submodular function is subadditive.
  Notice   that $I$ and $R$ are i.i.d. variables. This  along with linearity of expectation implies  
  \begin{align} \label{eq:submodUpper}
 \adap(\T,f) \leq \E_{I,R} [2 \cdot f(R)  +
  \adap(\T_I,f_{I\cup R})]. 
   \end{align}
  
Next, we lower  bound the expected value of the non-adaptive strategy from Eq.~\eqref{eq:expansionAdapNA}. We use monotonicity of $f$ to get
\begin{align} \alg(\T,f) \quad=  \quad\E_{I,R} [f(R) + \alg(\T_I,f_R)] \quad \geq \quad \E_{I,R} [f(R) + \alg(\T_I,f_{I \cup R})].  \label{eq:submodLower}
\end{align} 
 Since $f_{I \cup R}$ is  also a monotone submodular
function over independent elements and $\T_I$ is an adaptive strategy tree of depth at most $d-1$, by
induction hypothesis  
\[ \alg(\T_I,f_{I \cup R}) \geq \frac12
\adap(\T_I,f_{I\cup R}).\] 
Combining this with Eq.~\eqref{eq:submodUpper} and Eq.~\eqref{eq:submodLower}, we get
\[ \alg(\T,f) \geq \frac12
\adap(\T,f),\] 
which finishes the proof of the upper bound by induction.
\end{proof}
\subsection{Lower Bound of $2$} \label{sub_lower_bound}

In this section  we show  a monotone non-negative submodular function and a prefix-closed set of constraints where the adaptivity gap for stochastic probing is arbitrarily close to $2$.
Combined with \S\ref{sec:submodUpper},  this proves Theorem~\ref{thm:SubmodAdapGap} that the optimal adaptivity gap is exactly $2$. 
 

The proof below uses a stochastic probing instance on an infinite universe. Since submodular functions are  defined only on  finite sets, the proof below is informal. We do this to explain our main ideas and defer the formal proof to \Cref{sec:formalAdapLower}.

\begin{proof}[Informal proof of the lower bound in \Cref{thm:SubmodAdapGap}] 
 Our example is on a universe $V := \{e_{(k, l)} \mid k, l \in \integers_{\geq 0} \}$ 
    where every element is independently active with probability $\epsilon$ for some $0 <
    \epsilon < 1$. 
    
   \paragraph{Example:} We define our submodular objective $f$ to be the  weighted rank 
   function     of a  partition matroid that selects at most one element from each part.
    The elements are partitioned according to their first
    label---for every $k \in \integers_{\geq 0}$ the set $\{e_{(k, l)} \mid l \in
    \integers_{\geq 0}\}$ is a part of the partition matroid with  weight
    $(1-\epsilon)^k$. In other words,  for any set $S
    \subseteq V$ let $K(S) := \{ k \mid e_{(k, l)} \in S \}$ be the (unique)
    set of first labels, then 
    \[f(S) \defeq \sum_{k \in K(S)} (1-\epsilon)^k. \]
    Note that this series always converges so $f$ is well defined.
    
   To define our prefix-closed constraints, 
    we consider an infinite 
    directed acyclic graph where every element is
    identified with a single node in the graph.  Every node/element
    $e_{(k,l)}$ has exactly two outgoing edges: towards $e_{(k,l+1)}$ and
    towards $e_{(k+l+1,0)}$. We denote $\{ e_{(k,0)},
    e_{(k,1)}, \ldots \}$ as the elements on  \emph{column} $k$. The
    probing constraint is that a sequence of elements can be probed if and
    only if it corresponds to a directed path starting at $e_{(0,0)}$. 
    See \Cref{fig:submodLower}     for an illustration.


\captionsetup{singlelinecheck=false}
\tikzstyle{graphnode}=[circle, draw, fill=yellow!70, inner sep=0pt, minimum width=16pt]
\begin{figure}[ht]
\begin{center}
\begin{tikzpicture}[thin,scale=1.3]

	\foreach \y in {0,1,2}{
		\foreach \x in {0, ..., \y}{
		\draw [->-,line width=0.2mm] (\x,-\y) -- (\x,-\y-1);
		}
	}

		\draw [->-,line width=0.2mm] (0,0) -- (1,-1);

		\draw [->-,line width=0.2mm] (0,-1) -- (2,-2);
		\draw [->-,line width=0.2mm] (1,-1) -- (2,-2);

		\draw [->-,line width=0.2mm] (0,-2) -- (3,-3);
		\draw [->-,line width=0.2mm] (1,-2) -- (3,-3);
		\draw [->-,line width=0.2mm] (2,-2) -- (3,-3);

		
	\foreach \y in {0,1,2,3}{
		\foreach \x in {0, ..., \y}{
			\node at (\x,-\y) [graphnode]{\x,\pgfmathparse{\y-\x} \pgfmathprintnumber{\pgfmathresult}};
		}
	}
\end{tikzpicture}
\end{center}
\vspace{-0.5cm}
\caption{Adaptivity gap lower bound  example for monotone submodular functions.}
\label{fig:submodLower}
\end{figure}
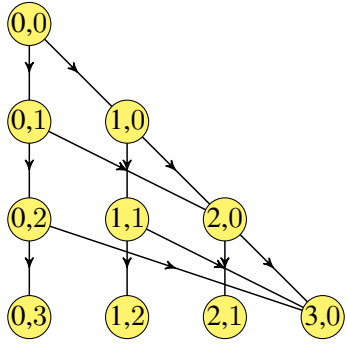

   
\paragraph{Analysis:}  
    We first give an adaptive strategy with value $2-\epsilon$ (in Eq.~\eqref{eq:lowerBoundSubmodAdap}) and later argue that 
    every non-adaptive strategy has value at most $1$ (in Eq.~\eqref{eq:lowerBoundSubmodAlg}); thereby, proving this theorem.
    Although, the probing constraint allows for infinite strategies, and in a different
    setting it would not be clear how to define their expected values,
     since $f$ is monotone we include every active element  in
    the solution.  So the expected value of an infinite strategy can be defined
    as the limit of strategies that only probe a finite number of
    elements. The finite lower bound example  in \Cref{sec:formalAdapLower} is
    constructed by reducing $V$ so that the resulting strategies are close
     to this limit.

    Our adaptive strategy \adap starts with probing element $e_{(0, 0)}$. 
    It is defined recursively:  after probing $e_{(k, l)}$,
    the next element to probe is either $e_{(k+l+1, 0)}$ if $e_{(k, l)}$ is found 
    active, or $e_{(k, l+1)}$ otherwise. In other words, it probes 
    elements on a column until it finds one active, and then probes
    another column.
    
    Let $\adap(k)$ denote the expected additional value our above adaptive strategy 
    if the next probed element is $e_{(k, 0)}$ and let $\adap \defeq \adap(0)$ denote
    the expected value of the entire strategy. Note that $\adap(k)$ does not
    depend on the set of elements found  active before probing $e_{(k,
    0)}$ (i.e.,  the elements $e_{(k', l')}$ where $k' < k$).  Furthermore,
    the subgraph reachable from $e_{(k, 0)}$ is similar to the entire graph on $\univ$
    in the sense that one can relabel the elements in the subgraph to match
    the entire graph exactly, the only difference being that the value of any
    subset is multiplied by a factor of $(1-\epsilon)^k$.  Therefore, we have
    \[\adap(k) = (1-\epsilon)^k \cdot \adap(0). \] 
    Now, summing over the number of
    inactive elements on  column $0$, we get
    \begin{align*} \adap(0) \quad = \quad \sum_{k=0}^{\infty} (1-\epsilon)^k \cdot \epsilon \cdot \Big(1 +    \adap(k+1) \Big) \quad = \quad \sum_{k=0}^{\infty}  (1-\epsilon)^k \cdot \epsilon \Big(1 +
  (1-\epsilon)^{k+1} \cdot \adap(0) \Big),
\end{align*} 
    which uses $\adap(k) = (1-\epsilon)^k \cdot \adap(0)$.
    Solving this equation yields the result: 
    \begin{align} \label{eq:lowerBoundSubmodAdap}
 \adap = \adap(0) = 2-\epsilon. 
     \end{align}
    
    Similarly, let $\alg(k)$ denote the expected additional value of the optimal
    non-adaptive strategy if the next probed element is $e_{(k, 0)}$, and let
    $\alg = \alg(0)$ denote the expected value of the optimal non-adaptive strategy. By the same argument as $\adap(k)$, we have 
    \[\alg(k) = (1-\epsilon)^k \cdot \alg (0). \] 
    Let $k$ denote the number of elements the optimal non-adaptive
    strategy probes on column $0$. 
    We get 
    \begin{align*} 
    \alg(0) \quad= \quad \sup_{k \geq 1} \Big\{1 - (1-\epsilon)^k + \alg(k) \Big\} \quad = \quad \sup_{k \geq 1} \Big\{1 -
(1-\epsilon)^k + (1-\epsilon)^k \cdot \alg(0)\Big\}, 
\end{align*}
  which uses $\alg(k) = (1-\epsilon)^k \cdot \alg(0)$. This implies 
  \begin{align} \label{eq:lowerBoundSubmodAlg}
  \alg = \alg(0) =  1.
  \end{align}
    
    Combining Eq.~\eqref{eq:lowerBoundSubmodAdap} and Eq.~\eqref{eq:lowerBoundSubmodAlg},  we get an adaptivity gap arbitrarily close to $2$ for $\epsilon
  \rightarrow 0$.  \end{proof}




\section{Adaptivity Gaps for a Weighted Rank Function of a $k$-Extendible
System}\label{sec:kSystem} For a downward-closed family $\calF$, recollect
that we define its rank function $f_\calF : 2^\univ \rightarrow \reals_{\geq
0}$ to be the largest cardinality subset in $\calF$, i.e., $f_\calF(S) \defeq
\max_{T \subseteq S ~\&~ T \in \calF} |T| = \max_{T \in \calF} |S \cap T|$.
In this section we prove our results on the adaptivity gaps of a
weighted rank function of  a $k$-extendible system.  \kSystemMain*

In \S\ref{sec:unwtdKSystem} we prove the upper bound for unweighted
$k$-extendible systems, and in \S\ref{sec:wtdKSystem} we give a reduction
from weighted to unweighted $k$-extendible systems that loses a factor
$O(\log k)$ in the adaptivity gap.  Our lower bound is presented in
\S\ref{sec:lowerKSystem}.

To simplify our proofs, we define an element $e \in T$ as a \emph{loop} in
$\calF \subseteq 2^T$ if $\{ e \} \not \in \calF$. Furthermore, given
a non-loop element $e \in T$, we define the \emph{contraction} $\calF / e$  as
$\{ F \setminus \{ e \} \mid F \in \calF, e \in F \}$, i.e., the family of
subsets that contain $e$ but with $e$ removed. We also need the following  property of $k$-extendible systems, which intuitively means a set $E \in \calF$ hurts at most $k \cdot |E|$ from another set $B \in \calF$. We include the proof for completeness in \Cref{appendix:k-extendible-property-for-sets-proof}.

\begin{restatable}{fact}{kExtendiblePropertySets} 
  \label{kExtendiblePropertySets}
  Let $\calF \subseteq 2^T$ be a $k$-extendible system. For every $A \subseteq B \in \calF$ and $E \subseteq T$ where $A \cup E \in \calF$, there exists a set $Z \subseteq B \setminus A$ such that $|Z| \le k \cdot |E|$ and $B \setminus Z \cup E \in \calF$.
\end{restatable}

\subsection{Upper Bound of $2k$ for an Unweighted $k$-Extendible System}
\label{sec:unwtdKSystem} Let $\T$ denote the optimal adaptive strategy for
maximizing the rank function $f$ of a given $k$-extendible system $\calF$. We
prove the following unweighted upper bound of \Cref{thm:kSystemAdapGap}.

\begin{theorem}\label{thm:unwtdKSystem} The adaptivity gap for \SMP
where the constraints are prefix-closed and the function is an unweighted
rank function of a $k$-extendible system is at most $2k$.  \end{theorem}

We use the random walk strategy to convert the adaptive strategy $\T$ into a
non-adaptive strategy. To analyze our algorithm, we define a natural \emph{greedy
procedure} to select a  subset of $A \subseteq T$ that is also in $\calF
\subseteq 2^T$.  First,  consider elements of $A$ in an arbitrary order (which
can even be determined on the fly). If the currently considered element is a
non-loop, it gets contracted in $\calF$; otherwise it gets ignored. Any such
computed set is in $\calF$ and the final output, the number of contracted
elements, is denoted by $\greedy(A)$. We first show that for $k$-extendible
systems such a greedy procedure produces a $k$-approximation to the largest
subset in $\calF$. A similar statement has been proven by
Mestre~\cite{mestre2006greedy}. 

\begin{lemma} \label{lemma:greedy} Let $f$ be a rank function of a
  $k$-extendible system $\calF \subseteq 2^T$. Fix any subset $A \subseteq T$
  and consider the output of the greedy procedure $\greedy(A)$ with an
  arbitrary ordering of $A$. We have that $f(A) \le k \cdot \greedy(A)$.
Even more, for any $A \subseteq B \subseteq T$ we have that $f(A) \le k \cdot
greedy(B)$.  \end{lemma} 

\begin{proof}
  Let $G \subseteq B$ be the set picked by $\greedy(B)$. Notice that $G$ is a maximal set in $\calF$ (need not be maximum). On the other hand, let $\Opt \subseteq A$ be the set picked by $f(A)$, i.e., the maximum set in $\calF$ on $A$. Our goal is to prove $|\Opt| \le k \cdot |G|$.

  Let $C \defeq \Opt \cap G$, note that $G = C \cup (G \setminus C) \in \calF$ and $C \subseteq \Opt$, hence by \Cref{kExtendiblePropertySets} there is a $Z \subseteq \Opt \setminus C$ with $|Z| \le k \cdot |G \setminus C| = k\cdot |G| - k\cdot  |C|$ such that $\Opt \setminus Z \cup (G \setminus C) = (\Opt \setminus C) \setminus Z \cup G \in \calF$. However, since $G$ is a maximal set and $(\Opt \setminus C) \cap G = \emptyset$ we know that $\Opt \setminus C \setminus Z = \emptyset$ and hence $|\Opt| \le |Z| + |C| \le k \cdot |G| - k \cdot |C| + |C| = k \cdot |G| - (k-1)|C| \le k \cdot |G|$.
\end{proof}

Given the above properties of a $k$-extendible system, we can now prove
\Cref{thm:unwtdKSystem}.

\begin{proof}[Proof of \Cref{thm:unwtdKSystem}] 
Let $\vX$ and $\vX'$ denote the element types for the adaptive and the non-adaptive algorithms, respectively. The adaptive strategy on the
optimal decision tree $\T$ gets  value $f(\vX_S)$, where $S \subseteq \univ$ is the set of probed elements by strategy $\T$ for type vector $\vX$.  We compare this value to a {greedy} strategy $\greedy(\vX_S \cup   \vX'_S)$ in which
\begin{enumerate}[itemsep=0ex,label=(\alph*),topsep=0pt,parsep=0pt] \item we
  consider the elements of $S$ in root-to-leaf order in which they
  appear on the tree and 
\item for any  $e \in S$  we first consider
  $\vX'_e$ (the true type) before $\vX_e$ (the virtual type) in the greedy
order.  \end{enumerate} Note by \Cref{lemma:greedy} we have  
\[ \adap(\T,f) = \E_{\vX}[ f(\vX_S) ]
  \le k \cdot \E_{\vX, \vX'} [ \greedy(\vX_S \cup \vX'_S)] . \] 
  By induction on the subtrees,
below we  prove  
\begin{align} \label{eq:GreedyVsAlg} 
\E_{\vX, \vX'} [\greedy(\vX_S \cup \vX'_S)] \leq  2\cdot \alg(\T,f).  \end{align} This finishes the proof of
\Cref{thm:unwtdKSystem} because the optimal non-adaptive algorithm has value
at least \[	\alg(\T,f) \geq \frac12 \cdot  \E_{\vX, \vX'} [\greedy(\vX_S \cup \vX'_S)]  \geq
\frac{1}{2k} \cdot  \adap(\T,f).  \]
 
To prove the missing Eq.~\eqref{eq:GreedyVsAlg}, we induct on the height of
the tree and $\calF$ being any downward-closed family.  For consistency, we
define the notation of $\greedy(\T, f)$ to denote the value of the above
greedy strategy when run on $\T$ with a rank function $f$. Thus,
 $\greedy(\T, f) = \E_{\vX, \vX'} [\greedy(\vX_S \cup \vX'_S)]$.
Suppose $e \in
\univ$ is the label of the root of $\T$. Denote by $I \defeq X_e$ the (random)
type of element $e$ when probed by the adaptive strategy (which is also the
virtual type of the non-adaptive strategy), and denote $R \defeq X'_e$  the
(random) true type when probed by the non-adaptive strategy. Also, let $\T_I$
denote the subtree the adaptive strategy goes to when the root $e$ is in
state $I$. We have \begin{align*} \greedy(\T,f) \leq \E_{I,R} [f(I \cup R) +
\greedy(\T_I, (f / R) / I)] , \end{align*} where by $(f / R) / I$ we mean the
rank function of $\calF$ after we first contract $R$  if it a
non-loop, and then contract $I$  if it is still a non-loop.  Now
subadditivity of $f$ gives \begin{align} \greedy(\T,f) & \leq \E_{I,R} [f(I)
+ f(R) + \greedy(\T_I, (f / R) / I)] \notag \\ & = \E_{I,R} [2 \cdot f(R) +
\greedy(\T_I, (f / R) / I)],  \label{eq:kSystemUpper} \end{align} where the
last equality uses linearity of expectation as $I$ and $R$ are identically
distributed.

Next, we lower bound the  value of our non-adaptive algorithm. Although it
takes a random root-leaf path and decides the set of elements to retain in
the end, we lower bound its value by an online algorithm that greedily
selects $R$ (unless it is a loop), however, always also contracts $I$ if it
is a non-loop. This gives, \begin{align} \label{eq:kSystemLower} \alg(\T,f)
&\geq \E_{I,R} [f(R) + \alg(\T_I, (f / R) / I)].  \end{align} Since $(f / R) /
  I$ is also a rank function of a downward-closed system and $\T_I$ is an
  adaptive strategy, by induction hypothesis we have \begin{align*}
  \alg(\T_I, (f / R) / I) \geq \frac{1}{2} \greedy(\T_I, (f / R) / I).
\end{align*} Combining this with Eq.~\eqref{eq:kSystemUpper} and
Eq.~\eqref{eq:kSystemLower}, we get \[ \greedy(\T,f) \leq 2 \cdot \alg(\T,f),
\] which proves Eq.~\eqref{eq:GreedyVsAlg} by induction.  \end{proof}

\subsection{Reducing Weighted  to Unweighted  $k$-Extendible System by Losing
$O(\log k)$} \label{sec:wtdKSystem}

We show how to extend the adaptivity gap result for an unweighted
$k$-extendible system to a weighted $k$-extendible system by losing an
$O(\log k)$ factor.  \begin{theorem}\label{thm:wtdKSystem}
  For \SMP over prefix-closed constraints, the adaptivity gap for a weighted
  rank function of a $k$-extendible system is at most $32 k \log_2 k$.
  \end{theorem} \begin{proof} Given a weighted rank function $f$ of a
  $k$-extendible system $\calF \subseteq 2^T$ over a set of types $T$, we
define $f_j$ for $j \in \integers$ to be an unweighted rank function of the
$k$-extendible system $\calF$; however, the new weights are changed such that
only the types with original weights in $( 2^{j-1},2^{j}]$ participate
with new weight of $1$, while the other elements have a new weight of $0$.
Note that this partitions the set of types $T$ into pairwise disjoint
\emph{classes}. Notice, we have 
\begin{align} \label{eq:adapWtdUpper} 
\adap(\T, f)  \leq
\sum_j 2^j \cdot \adap(\T, f_j),  \end{align}
 where $adap(\T, f_j)$ denotes
the expected value of an adaptive strategy given by the common decision tree
$\T$ with respect to the rank function $f_j$.  

Now, since $\adap(\T, f_j)$ is
an unweighted $k$-extendible system problem, we know  that a random root-leaf
path returns a solution with expected value \begin{align}
\label{eq:algUnwtdLower} \alg(\T,f_j) \geq \frac{1}{2k} \cdot \adap(\T, f_j).
\end{align} In the following lemma, we show that these non-adaptive solutions
for $f_j$ can be combined to obtain a feasible and ``high-value" non-adaptive
solution for $f$.  
\begin{lemma} \label{lem:nonAdapWtdLower} The random-walk non-adaptive
algorithm $\alg$ has expected value \[ \alg(\T,f) \geq \frac{1}{16\cdot \log
k}\sum_j 2^j \cdot \alg(\T,f_j).  \] \end{lemma} 
Before proving
\Cref{lem:nonAdapWtdLower}, we  finish the proof of \Cref{thm:wtdKSystem} by
combining it with Eq.~\eqref{eq:algUnwtdLower} and
Eq.~\eqref{eq:adapWtdUpper}: \begin{align*} \alg(\T,f) \geq \frac{1}{16\cdot
  \log k}\sum_j 2^j \cdot \alg(\T,f_j) &\geq \frac{1}{32k \log k}\sum_j 2^j
  \cdot \adap(\T,f_j) \geq \frac{1}{32k \log k} \cdot \adap(\T,f).	\qedhere
  \end{align*} \end{proof}

Informally, in the proof of \Cref{lem:nonAdapWtdLower} we combine the
unweighted solutions of $\alg(\T, f_i)$ by running a ``greedy-optimal" algorithm from
the higher weight to the smaller weight classes and fixing the types chosen
in earlier classes. Unfortunately, in general such an approach loses an
extra factor $k$ in the approximation. To fix this, our second idea is
to increase the weight gap between successive classes. We achieve this by
combining $O(\log k)$ consecutive classes into a \emph{bucket}, where in each
bucket we focus on the class with the largest non-adaptive value. 
Because of  boundary issues, we only take either odd or even buckets.


\begin{proof} [Proof of \Cref{lem:nonAdapWtdLower}]

Let $a \leq b \in \integers$ denote the indices of the smallest and the
highest weight classes. We define buckets consisting of $2\log k$ consecutive
classes, where bucket $B_i$ consists of classes $\{ b - 2i  \log k, b - 2i  \log k-1, \ldots,
b - 2(i-1) \log k \}$. For each $B_i$, let \[j(i) \defeq \argmax_{j \in
  B_i}\{2^j \cdot \alg(\T,f_j)  \}. \] Since each bucket has size $2\log k$,
  this implies \[  \sum_{i } 2^{j(i)} \cdot \alg(\T,f_{j(i)}) \geq
  \frac{1}{2\cdot \log k} \sum_j 2^{j} \cdot \alg(\T,f_{j}).  \] Without loss
  of generality we can assume the odd indices satisfy \[ \sum_{i \text{ is
  odd}} 2^{j(i)} \cdot \alg(\T,f_{j(i)}) \geq \frac12  \sum_i 2^{j(i)} \cdot
\alg(\T,f_{j(i)}).  \] Otherwise, use the same argument for even indices.
Combining the last two equations, we get 
\begin{align} \label{eq:oddToAll}
\sum_{i \text{ is odd}} 2^{j(i)} \cdot \alg(\T,f_{j(i)}) \geq
\frac{1}{4\cdot \log k} \sum_j 2^{j} \cdot \alg(\T,f_{j}).  
\end{align} 

We now claim that a  \emph{greedy-optimal} algorithm has a large value: It goes  
over classes $j(i)$ in decreasing order of (odd) buckets, but  it always selects the
maximum independent set (instead of selecting a  maximal greedy set) in the 
current class $j(i)$ given its choices in the previous. 
This algorithm is, therefore, a combination of greedy and optimal algorithms.

\begin{claim} \label{claim:greedyOdd} 
Consider an   algorithm that goes over the {odd numbered} buckets in decreasing order of
  weights and selects the \emph{maximum} set from class $j(i)$ in bucket $i$ 
  such   that the resulting set is still feasible in $\calF$. (After a set in a class is selected,
   it gets   fixed for all smaller choices.) The finally chosen set  has value at least
\begin{align*} \frac{1}{4} \sum_{i \text{ is odd}} 2^{j(i)} \cdot
\alg(\T,f_{j(i)}) .  \end{align*} \end{claim}

\begin{proof} The intuition  is that for a $k$-extendible system by 
\Cref{kExtendiblePropertySets} any selected
  member can ``hurt'' at most $k$ members from lower buckets.  Since we only
  consider odd numbered buckets, two types in different buckets differ in
  their weights by at least a factor of $2^{2 \log k} = k^2$.  Thus, losing
  $k$ types of lower weight should not significantly impact the value.

Let $\ell$ be the random variable denoting the leaf reached by the random walk
on the decision tree $\T$, and let $R$ be the random set of elements seen
 by the random-walk non-adaptive strategy on this path. Furthermore, let $A_i$
denote the set of elements picked by the non-adaptive strategy with 
respect to $f_{j(i)}$,  let $A'_i \subseteq A_i$ be the set of elements picked
by our greedy-optimal non-adaptive strategy from bucket $i$, and let 
$A'_{<i} $ denote $\bigcup_{i'<i ~:~ i' \text{ is odd}} A_{i'}$. In other words, 
$A'_{<i}$ is the greedy-optimal solution up to bucket number $i$ and $A'_i$ is the
maximum  subset of $A_i$ such that $A'_i \cup A'_{<i} \in \calF$.  Note that
$A_i$, $A'_i$ and $A'_{<i}$ are random variables depending on $\ell$ and $R$.


Using \Cref{kExtendiblePropertySets} on the $k$-extendible system $\calF$ with the preconditions $\emptyset \cup A'_{< i} \in \calF$ and $\emptyset \subseteq A_i$,
 there exists a set $Z$ with $|Z| \le k \cdot |A'_{< i}|$ such that $A_i \setminus Z \in \calF$. Hence, we have
\begin{align*} 
|A'_i| \ge |A_i \setminus Z| &\geq |A_i| - k \cdot |A'_{<i}|.  
\end{align*}
  Multiplying by $2^{j(i)}$ and summing over all odd $i$ gives 
  \begin{align}
    \label{eq:chargingKSystem} \sum_{i \text{ is odd}} 2^{j(i)} \cdot |A'_i|
    &  \geq \sum_{i \text{ is odd}} 2^{j(i)} \cdot |A_i| - k\cdot \sum_{i
  \text{ is odd}} 2^{j(i)} \cdot |A'_{<i}| \notag \\ &  = \sum_{i\text{ is
odd}} 2^{j(i)} \cdot |A_i| - k \cdot \sum_{i\text{ is odd}} |A'_i| \sum_{i'>i
~:~ i' \text{ is odd}} 2^{j(i')}.  
\end{align} 
Now, since every bucket $i$
contains $2\log k$ classes, where two successive class weights differ by a
factor of $2$, we know \[2^{j(i+2)} \leq \frac{2^{j(i)}}{k^2}. \] Combining
this with Eq.~\eqref{eq:chargingKSystem} gives \begin{align*} \sum_{i \text{
  is odd}} 2^{j(i)} \cdot  |A'_i| &  \geq \sum_{i\text{ is odd}} 2^{j(i)}
      \cdot |A_i| - k \cdot \sum_{i\text{ is odd}} |A'_i| \sum_{i' >  i ~:~
    i' \text{ is odd}} \frac{2^{j(i'+2)}}{k^2}\\ &  \geq \sum_{i\text{ is
  odd}} 2^{j(i)} \cdot |A_i| - \sum_{i\text{ is odd}} |A'_i| \cdot
{2^{j(i)}}, \end{align*} where the last inequality uses \begin{align*}
  \sum_{i'>i~:~ i' \text{ is odd}} {2^{j(i'+2)}} \quad = \quad \sum_{i'\geq i
  ~:~ i' \text{ is odd}} {2^{j(i')}} \quad \leq \quad 2 \cdot 2^{j(i)} \quad
  \leq \quad k \cdot 2^{j(i)}.  \end{align*} After rearranging,
\begin{align*} \sum_{i \text{ is odd}} 2^{j(i)} \cdot  |A'_i|  \geq \frac12
\cdot  \sum_{i\text{ is odd}} 2^{j(i)} \cdot |A_i|  .  \end{align*}

Notice that by definition of a class, each type in class $j(i)$ has weight at
least $2^{j(i)-1}$. Using this fact and taking expectation over $\ell$ and $R$,
we get 
\begin{align*} 
\alg(\T, f) &\geq \E_{\ell, R} \Big[\sum_{i \text{ is odd}}
2^{j(i)-1} \cdot |A'_i| \Big] \\ 
&\geq \frac{1}{4} \E_{\ell, R} \Big[\sum_{i \text{ is odd}} 2^{j(i)} \cdot |A_i| \Big] = \frac{1}{4} \sum_{i \text{ is odd}} 2^{j(i)} \cdot \alg(\T, f_{j(i)}), 
\end{align*}
which finishes the proof of \Cref{claim:greedyOdd}. 
\end{proof}
Using
\Cref{claim:greedyOdd}, we have 
\begin{align*} 
\alg(\T,f) \geq  \frac14 \sum_{i \text{ is odd}} 2^{j(i)} \cdot \alg(\T,f_{j(i)}), 
\end{align*}  
which combined when with Eq.~\eqref{eq:oddToAll} proves \Cref{lem:nonAdapWtdLower}.  
\end{proof}

\subsection{Lower Bounds} \label{sec:lowerKSystem}

We present two very similar lower bound examples: one where the adaptivity
gap is $k-o(1)$ for a rank function of an unweighted $k$-extendible system
and another where the adaptivity gap is $\Omega(\sqrt{k})$ for a rank
function of an intersection of $k$ matroids. A related example was also shown
in~\cite{GNS-SODA17}.

\paragraph{Example:} For generality we work in the Bernoulli setting where each element in $\univ$ is either active or inactive. Consider a perfect $w$-ary tree of depth $k$ whose edges
correspond to the ground set $V$. Each edge is active with probability $p > 0$. For any leaf
$\ell$, let $P_\ell$ denote the unique path from the root to $\ell$.  The objective
value on any set is the maximum number of edges
in the set on the \emph{same} root-leaf path, i.e.,  for any $S
\subseteq V$, 
\[ f(S) \defeq \max_{\text{leaf }\ell} |P_\ell \cap S| . \]  

The feasibility constraints are such that a set of edges can be probed if and only if there exists some root-leaf path $P_\ell$ such that every probed edge has at least one endpoint on $P_\ell.$ Note that this implies that a maximum of $w \cdot k$ edges can be probed.

\paragraph{Analysis:}
Let the adaptive strategy be the following: probe all $w$ edges incident to
the root. If any of them is active, start probing the edges directly below
the active edge, otherwise below the first edge. Continue recursively
until a leaf is reached. On every level, the adaptive strategy has $1 - (1 -
p)^w$ probability of finding an active edge. Therefore, the expected value of
the adaptive strategy is $k \cdot (1 - (1 - p)^w).$

For any non-adaptive strategy, the feasibility constraints imply 
there exists a root-leaf path $P_\ell$ such that
all probed edges have an endpoint on it. Suppose all $w \cdot k$
 edges incident to $P_\ell$ are probed. The non-adaptive strategy can get value at most $1$
from the edges not on $P_\ell$ and in expectation at most $k \cdot p$ from the
edges on $P_\ell.$ So, the non-adaptive strategy has an expected value of at
most $1 + k \cdot p.$

\subsubsection*{Lower Bound of $k$ for an unweighted $k$-extendible system}

Consider the example described above and set $w \defeq k^4$ and $p \defeq \frac{1}{k^3}$.
The function $f$ is trivially a rank function of a $k$-extendible system because the rank of the system is $k$, i.e., $f(\univ) = k$.
The adaptive strategy has an expected value
\[ k \cdot \Big(1 - \Big(1 - \frac{1}{k^3}\Big)^{k^4} \Big) \geq k \cdot \Big(1 - \frac{1}{e^k} \Big) = k - o(1), \] 
whereas any non-adaptive strategy has an expected value at most 
$1 + \frac{1}{k^2}.$
This gives an adaptivity gap of $k - o(1)$.

\subsubsection*{Lower Bound of $\Omega(\sqrt{k})$ for an unweighted
intersection of $k$ matroids} 

In this section we show how to model the above example as an intersection of $t = k^2$ matroids, yielding an adaptivity gap of $\Omega(\sqrt{t})$ for an intersection of $t$ matroids. Consider the example described above and set $w \defeq k$ and $p \defeq \frac{1}{k}$.
The adaptive strategy has an expected value of
\[ k \cdot \Big(1 - \Big(1-\frac{1}{k} \Big)^k \Big) \geq k \cdot \Big(1 - \frac{1}{e} \Big) = \Omega(k) \] 
and the non-adaptive strategy gets at most $2$ in expectation; so the adaptivity gap is $\Omega(k)$.

All that remains to show is that $f$ can be represented as an intersection of $k^2$ simple partition matroids. We use the term simple partition matroid for a matroid that partitions the $\univ$ into multiple parts and a set is independent if it contains at most one element in every part.

Suppose that $k$ is prime and label each node $v$ with a list $L_v$ as follows: the root's label is an empty list $()$.
Let $L(i)$ denote the $i^{th}$ element of the list $L$
and $L + x$ a list equal to $L$ with $x$ appended to it. All the other nodes
are labeled recursively: let $v$ be a node with children $\{v_0, v_1, ...
v_{k-1}\}$. Define $L_{v_i} \defeq L_v + i$. Hence, $u$ is an ancestor of $v$ if
and only if $L_u$ is a prefix of $L_v$, and otherwise $L_u(i) \neq L_v(i)$ for
some $i$.
    
    Let $e_v$ denote the edge/element between $v$ and its parent. We 
    define $k^2$ partition matroids $M_{i, j}$ for $i \in \{1, 2, ..., k\}$ and  $j
    \in \{0, 1, ..., k-1\}$. Each $M_{i, j}$  consists of $k$ \textit{big}
    partitions indexed from $0$ to $k-1$, and all other partitions  contain
    only a single element. Let 
    \[ I_v(i, j) \defeq L_v(i)j + d_v (\mathrm{mod}\ k). \]
    For a node $v$ on depth $d_v \geq i$, element $e_v$ is in
    the $I_v(i, j)^{th}$ \textit{big} partition of $M_{i, j}$. For a node $v$ on depth $d_v
    < i$, $e_v$ is the only element in its partition in $M_{i, j}$.
    
    We claim that $f$ is the rank function of $\calF \defeq \bigcap_{i=1}^{k}
    \bigcap_{j=0}^{k-1} M_{i, j}$, which is an  intersection of  $k^2$ matroids.
    Since $\calF$ is an intersection of simple partition matroids,
    $S \in \calF$ if and only if $\{a, b\} \in \calF$ for every $a, b \in S$.
    Now consider two nodes $u, v$ such that $\{e_u, e_v\} \not\in \calF$. This means
    $I_u(i, j) = I_v(i, j)$ for some $i \leq d_u, d_v$ and $j \in \{0, 1,
    ..., k-1\}$, which is equivalent to 
    \[ L_u(i) \cdot j + d_u \equiv L_v(i) \cdot j + d_v
    (\mathrm{mod}\ k). \]
     Since $k$ is prime, this holds for some $i, j$ if
    and only if $d_u = d_v$ (for $j = 0, i = 1$) or $L_u(i) \neq L_v(i)$ for
    any $i$. That is, $\{e_u, e_v\} \not\in \calF$ if and only if $u$ and $v$
    are not ancestors of one another, which completes the proof.




\medskip
\noindent
{\bf Acknowledgments}.
We thank  Anupam Gupta   for useful discussions.  The second author was supported in part by NSF awards CCF-1319811, CCF-1536002, and CCF-1617790. The third author was supported in part by CCF-1527110, CCF-1618280 and NSF CAREER award CCF-1750808.

\appendix

\section{Adaptivity Gap Lower Bound of $2$ for Submodular Functions} \label{sec:formalAdapLower}

\begin{proof}

  As mentioned, the finite lower bound example is constructed by reducing the
  infinite example given in Section \ref{sub_lower_bound}. However, this reduction
  loses the nice similarity properties of the graph so much more calculation is
  required in order to bound the strategies.

  Let $0 < \epsilon < 1/2$ and $D$ be the smallest integer such that
  $(1-\epsilon)^D < \epsilon^2$. The ground set is the result of removing
  elements $e_{(k,l)}$ where $k+l > D$, that is $V \defeq \{ e_{(k, l)} : k, l \in
  \mathbb{Z}_{\ge 0}, k + l \le D \}$ where each node is active with probability
  $\epsilon$.  The probing constraint and the objective function $f$ are
  naturally reduced to this set: a sequence of elements can be probed if they
  correspond to a (finite) path starting at $e_{(0,0)}$ in the given graph, and
  $f(S) \defeq \sum_{k \in K(S)} (1 - \epsilon)^k$ where $K(S)$ is the set of
  (unique) first labels which now finite.  Similarly as before, we will denote
  $\{ e_{(k,0)}, e_{(k,1)}, \ldots, e_{(k, D-k)} \}$ as the vertices on the
\textit{column $k$}.

  We first show that any non-adaptive strategy has expectation at most 1. Let
  $\alg(k)$ denote the additional expected value of the optimal non-adaptive strategy if the
  next probed element is $e_{(k, 0)}$. We will inductively prove $\alg(k) <
  (1-\epsilon)^k$, which is sufficient for our claim. For the base case $k = D$,
  the inequality clearly holds since $\alg(D) = \epsilon (1-\epsilon)^D <
  (1-\epsilon)^D$. For $0 \le k < D$ let $i$ be the second label of the last
  vertex probed on the column $k$.  
  \begin{align*} \alg(k) \quad &=
  \max_{i=0}^{D-k} \Big[(1-\epsilon)^k \Pr[k \in K(\text{active})] + \alg(k+i+1)\Big] \\
  &= \max_{i=0}^{D-k} \Big[(1-\epsilon)^k (1 - (1-\epsilon)^{i+1}) + \alg(k+i+1)\Big]
  \\
  &< \max_{i=0}^{D-k}\Big[(1-\epsilon)^k(1 - (1-\epsilon)^{i+1}) +
(1-\epsilon)^{k+i+1} \Big]  \quad = \quad (1-\epsilon)^k .
    \end{align*} 
    This completes the induction and proves that non-adaptive strategies get at most $1$.

Finally, we show that there exists an adaptive strategy with expected value at
least $2 - O(\epsilon)$ for sufficiently small $\epsilon > 0$. This finalizes
the proof since it implies a gap of $2$ by taking $\epsilon \to 0$. The
strategy is naturally reduced: first probe $e{(0,0)}$ and after probing some
$e_{(k,l)}$ terminate if $k+l=D$, otherwise probe $e_{(k+l+1,0)}$ if
$e_{(k,l)}$ is active and $e_{(k,l+1)}$ if not. Let $\adap(k)$ denote the expected value this
strategy gets when the next probed element is $e_{(k, 0)}$, for $0 \le k \le D$.
For convenience, define $\adap(D + i) \defeq 0$ for all $i \ge 1$.

We prove by induction that $\adap(k) >
\frac{4-6\epsilon}{2-\epsilon}(1-\epsilon)^k - 8\epsilon$, which is sufficient
to finalize the proof since then $\adap(0) > 2 - O(\epsilon)$. For $k$ large
enough that $\frac{4-6\epsilon}{2-\epsilon}(1-\epsilon)^k < 8\epsilon$, the
inequality clearly holds and presents our base case.  Otherwise,
$(1-\epsilon)^k \ge 8 \frac{2-\epsilon}{4 - 6 \epsilon} \epsilon > 4\epsilon$.
Let $i$ be the second label of the last vertex probed on the column $k$ and let
$A$ denote the set of active elements.  
\begin{align*} \adap(k) &=
\sum_{i=0}^{D-k} \Pr \Big[v_{(k, i)} \in A, v_{(k, 0)} \not \in A, \ldots, v_{(k,
i-1)} \not \in A  \Big] \left[(1-\epsilon)^k + \adap(k+i+1)\right] \\
&= \sum_{i=0}^{D-k}(1-\epsilon)^i \epsilon \left[(1-\epsilon)^k +
    \adap(k+i+1)\right] \\
    &= \epsilon \cdot \sum_{i=0}^{D-k} (1-\epsilon)^{k+i} + \epsilon \cdot \sum_{i=0}^{D-k}
    (1-\epsilon)^i \adap(k+i+1) \\
    &= \epsilon \cdot \frac{1}{\epsilon}(1-\epsilon)^k \Big(1-(1-\epsilon)^{D-k+1} \Big) +
    \epsilon \cdot \sum_{i=0}^{D-k}(1-\epsilon)^i \cdot \adap(k+i+1).
  \end{align*}
  Using the induction hypothesis, we get
  \begin{align*}
    \adap(k) &> (1-\epsilon)^k - (1-\epsilon)^{D+1} + \epsilon
    \sum_{i=0}^{D-k}(1-\epsilon)^i \Big(\frac{4-6\epsilon}{2-\epsilon}(1-\epsilon)^{k+i+1}
    - 8\epsilon \Big) \\
    &= (1-\epsilon)^k - (1-\epsilon)^{D+1} + \epsilon \sum_{i=0}^{D-k}
    \frac{4-6\epsilon}{2-\epsilon}(1-\epsilon)^{k+2i+1} 
    - 8\epsilon^2 \sum_{i=0}^{D-k}(1-\epsilon)^i \\
    &= (1-\epsilon)^k - (1-\epsilon)^{D+1} + (1-\epsilon)^{k+1}
    \frac{4-6\epsilon}{(2-\epsilon)^2} \Big(1 - (1 - \epsilon)^{2(D-k+1)} \Big) \\
    &\phantom{=}\qquad - 8\epsilon \Big(1-(1-\epsilon)^{D-k+1} \Big).
  \end{align*}
  After dropping some positive summands and using $(1-\epsilon)^D < \epsilon$ and $(1-\epsilon)^k > \epsilon$, we get
  \begin{align*}
\adap(k) 
    &> (1-\epsilon)^k - \epsilon^2 + (1-\epsilon)^{k+1}
  \frac{4-6\epsilon}{(2-\epsilon)^2}(1 - \epsilon^2) - 8\epsilon .
  \end{align*}
  It is sufficient to prove \begin{align*} & ( 1-\epsilon)^k - \epsilon^2 -
  8\epsilon + (1-\epsilon)^{k+1} \frac{4-6\epsilon}{(2-\epsilon)^2}(1 -
  \epsilon^2) > \frac{4-6\epsilon}{2-\epsilon}(1-\epsilon)^k - 8\epsilon.
\end{align*} 
Multiplying by $\frac{(2-\epsilon)^2}{(1-\epsilon)^k} > 0$, we get
an equivalent statement to prove:
\begin{align*}  (2-\epsilon)^2 -
\epsilon^2 \cdot \frac{(2-\epsilon)^2}{(1-\epsilon)^k} +
(1-\epsilon)(4-6\epsilon)(1-\epsilon^2) > (4-6\epsilon)(2-\epsilon).
\end{align*} 
Finally, using $\epsilon^2 \frac{(2-\epsilon)^2}{(1-\epsilon)^k} <
\epsilon^2 (2-\epsilon)^2 \frac{1}{4\epsilon} = \epsilon + O(\epsilon^2)$ and
expanding out, we note that the left-hand side is $8 - 15\epsilon +
O(\epsilon^2)$, while the right-hand side is $8 - 16\epsilon + O(\epsilon^2)$.
Therefore, the inequality holds for sufficiently small $\epsilon > 0$.  This
concludes the proof.  \end{proof}

\section{Proof of the $k$-Extendible Property for Set Extension}
\label{appendix:k-extendible-property-for-sets-proof}
\kExtendiblePropertySets*
\begin{proof}
  Enumerate the elements $E = \{e_1, \ldots, e_r\}$ where $r \defeq |E|$ and denote by $E_i \defeq \{ e_1, \ldots, e_i \}$ for $0 \le i \le r$. Initialize $Z_0 \defeq \emptyset$ and consider the following procedure to construct $Z_1, Z_2, \ldots, Z_r$ that satisfies the invariants $A \subseteq B \setminus Z_i$, $B \setminus Z_i \cup E_i \in \calF$ and $|Z_i| \le k \cdot i$.

  In the $i^{th}$ step we have that $A \cup E_{i-1} \cup \{e_i\} \in \calF$ by downward-closeness and $A \cup E_{i-1} \subseteq B \setminus Z_{i-1} \cup E_{i-1}$ by the induction hypothesis. Hence by $k$-extendibility we can find $Z' \subseteq B \setminus (Z_{i-1} \cup A \cup E_{i-1})$ with $|Z'| \le k$ and where $(B \setminus Z_{i-1} \cup E_{i-1}) \setminus Z' \cup \{ e_i \} = B \setminus (Z_{i-1} \cup Z') \cup E_i \in \calF$. Set $Z_i \defeq Z_{i-1} \cup Z'$ and note that $|Z_i| \le |Z_{i-1}| + |Z'| \le (i-1) \cdot k + k = i \cdot k$. Furthermore, already deduced that $B \setminus Z_i \cup E_i \in \calF$ and finally $A \subseteq B \setminus Z_i = B \setminus Z_{i-1} \setminus Z'$ since $Z' \cap A = \emptyset$. We satisfied all stipulations of the induction, hence we report $Z_r$ as the solution.
\end{proof}


\bibliographystyle{alpha}
\bibliography{bib}

\newcommand{\etalchar}[1]{$^{#1}$}
\begin{thebibliography}{GKMR11}

\bibitem[Ada11]{Adamczyk-IPL11}
Marek Adamczyk.
\newblock Improved analysis of the greedy algorithm for stochastic matching.
\newblock {\em Inf. Process. Lett.}, 111(15):731--737, 2011.

\bibitem[AGM15]{AGM-ESA15}
Marek Adamczyk, Fabrizio Grandoni, and Joydeep Mukherjee.
\newblock Improved approximation algorithms for stochastic matching.
\newblock In {\em Algorithms-ESA 2015}, pages 1--12. Springer, 2015.

\bibitem[AN16]{AN16}
Arash Asadpour and Hamid Nazerzadeh.
\newblock Maximizing stochastic monotone submodular functions.
\newblock {\em Management Science}, 62(8):2374--2391, 2016.

\bibitem[ANS08]{ANS-WINE08}
Arash Asadpour, Hamid Nazerzadeh, and Amin Saberi.
\newblock Stochastic submodular maximization.
\newblock In {\em International Workshop on Internet and Network Economics},
  pages 477--489. Springer, 2008.
\newblock Full version appears as~\cite{AN16}.

\bibitem[AR12]{AR-AER12}
Itai Ashlagi and Alvin~E.\ Roth.
\newblock New challenges in multihospital kidney exchange.
\newblock {\em American Economic Review}, 102(3):354--59, 2012.

\bibitem[ASW14]{ASW14}
Marek Adamczyk, Maxim Sviridenko, and Justin Ward.
\newblock Submodular stochastic probing on matroids.
\newblock In {\em STACS}, pages 29--40, 2014.

\bibitem[BCN{\etalchar{+}}15]{BCNSX-APPROX15}
Alok Baveja, Amit Chavan, Andrei Nikiforov, Aravind Srinivasan, and Pan Xu.
\newblock Improved bounds in stochastic matching and optimization.
\newblock In {\em Approximation, Randomization, and Combinatorial Optimization.
  Algorithms and Techniques, {APPROX/RANDOM} 2015, August 24-26, 2015,
  Princeton, NJ, {USA}}, pages 124--134, 2015.

\bibitem[BGK11]{BGK-SODA11}
Anand Bhalgat, Ashish Goel, and Sanjeev Khanna.
\newblock Improved approximation results for stochastic knapsack problems.
\newblock In {\em SODA}, pages 1647--1665, 2011.

\bibitem[BGL{\etalchar{+}}12]{BGLMNR-Algorithmica12}
Nikhil Bansal, Anupam Gupta, Jian Li, Juli{\'a}n Mestre, Viswanath Nagarajan,
  and Atri Rudra.
\newblock {When LP Is the Cure for Your Matching Woes: Improved Bounds for
  Stochastic Matchings}.
\newblock {\em Algorithmica}, 63(4):733--762, 2012.

\bibitem[BN14]{BN-IPCO14}
Nikhil Bansal and Viswanath Nagarajan.
\newblock On the adaptivity gap of stochastic orienteering.
\newblock In {\em IPCO}, pages 114--125, 2014.

\bibitem[CCPV11]{CCPV-SICOMP11}
Gruia C{\u{a}}linescu, Chandra Chekuri, Martin P{\'{a}}l, and Jan
  Vondr{\'{a}}k.
\newblock Maximizing a monotone submodular function subject to a matroid
  constraint.
\newblock {\em {SIAM} J. Comput.}, 40(6):1740--1766, 2011.

\bibitem[CIK{\etalchar{+}}09]{CIKMR-ICALP09}
Ning Chen, Nicole Immorlica, Anna~R. Karlin, Mohammad Mahdian, and Atri Rudra.
\newblock Approximating matches made in heaven.
\newblock In {\em Automata, Languages and Programming, 36th International
  Colloquium, {ICALP} 2009, Rhodes, Greece, July 5-12, 2009, Proceedings, Part
  {I}}, pages 266--278, 2009.

\bibitem[CVZ14]{CVZ-SICOMP14}
Chandra Chekuri, Jan Vondr{\'{a}}k, and Rico Zenklusen.
\newblock Submodular function maximization via the multilinear relaxation and
  contention resolution schemes.
\newblock {\em {SIAM} J. Comput.}, 43(6):1831--1879, 2014.

\bibitem[DGV04]{DGV-FOCS04}
Brian~C. Dean, Michel~X. Goemans, and Jan Vondr{\'a}k.
\newblock Approximating the stochastic knapsack problem: The benefit of
  adaptivity.
\newblock In {\em Foundations of Computer Science, 2004. Proceedings. 45th
  Annual IEEE Symposium on}, pages 208--217. IEEE, 2004.

\bibitem[DGV05]{DGV-SODA05}
Brian~C. Dean, Michel~X. Goemans, and Jan Vondr{\'{a}}k.
\newblock Adaptivity and approximation for stochastic packing problems.
\newblock In {\em Proceedings of the Sixteenth Annual {ACM-SIAM} Symposium on
  Discrete Algorithms, {SODA} 2005, Vancouver, British Columbia, Canada,
  January 23-25, 2005}, pages 395--404, 2005.

\bibitem[DHK14]{DHK-SODA14}
Amol Deshpande, Lisa Hellerstein, and Devorah Kletenik.
\newblock Approximation algorithms for stochastic boolean function evaluation
  and stochastic submodular set cover.
\newblock In {\em Proceedings of the Twenty-Fifth Annual {ACM-SIAM} Symposium
  on Discrete Algorithms, {SODA} 2014, Portland, Oregon, USA, January 5-7,
  2014}, pages 1453--1466, 2014.

\bibitem[GKMR11]{GKMR-FOCS11}
Anupam Gupta, Ravishankar Krishnaswamy, Marco Molinaro, and R.~Ravi.
\newblock Approximation algorithms for correlated knapsacks and non-martingale
  bandits.
\newblock In {\em {IEEE} 52nd Annual Symposium on Foundations of Computer
  Science, {FOCS} 2011, Palm Springs, CA, USA, October 22-25, 2011}, pages
  827--836, 2011.

\bibitem[GKNR12]{GKNR-SODA12}
Anupam Gupta, Ravishankar Krishnaswamy, Viswanath Nagarajan, and R.~Ravi.
\newblock Approximation algorithms for stochastic orienteering.
\newblock In {\em Proceedings of the Twenty-Third Annual {ACM-SIAM} Symposium
  on Discrete Algorithms, {SODA} 2012, Kyoto, Japan, January 17-19, 2012},
  pages 1522--1538, 2012.

\bibitem[GM07]{GM-STOC07}
Sudipto Guha and Kamesh Munagala.
\newblock Approximation algorithms for budgeted learning problems.
\newblock In {\em STOC}, pages 104--113. 2007.
\newblock Full version as: \emph{Approximation Algorithms for Bayesian
  Multi-Armed Bandit Problems}, \url{http://arxiv.org/abs/1306.3525}.

\bibitem[GM09]{GM-ICALP09}
Sudipto Guha and Kamesh Munagala.
\newblock Multi-armed bandits with metric switching costs.
\newblock In {\em Automata, Languages and Programming, 36th Internatilonal
  Colloquium, {ICALP} 2009, Rhodes, Greece, July 5-12, 2009, Proceedings, Part
  {II}}, pages 496--507, 2009.

\bibitem[GN13]{GN-IPCO13}
Anupam Gupta and Viswanath Nagarajan.
\newblock A stochastic probing problem with applications.
\newblock In {\em Integer Programming and Combinatorial Optimization - 16th
  International Conference, {IPCO} 2013, Valpara{\'{\i}}so, Chile, March 18-20,
  2013. Proceedings}, pages 205--216, 2013.

\bibitem[GNS16]{GNS-SODA16}
Anupam Gupta, Viswanath Nagarajan, and Sahil Singla.
\newblock Algorithms and adaptivity gaps for stochastic probing.
\newblock In {\em Proceedings of the Twenty-Seventh Annual ACM-SIAM Symposium
  on Discrete Algorithms}, pages 1731--1747. SIAM, 2016.

\bibitem[GNS17]{GNS-SODA17}
Anupam Gupta, Viswanath Nagarajan, and Sahil Singla.
\newblock {Adaptivity Gaps for Stochastic Probing: Submodular and XOS
  Functions}.
\newblock In {\em Proceedings of the Twenty-Eighth Annual ACM-SIAM Symposium on
  Discrete Algorithms}, pages 1688--1702. SIAM, 2017.

\bibitem[HFX18]{FLX-ICALP18}
Jian~Li Hao~Fu and Pan Xu.
\newblock {A PTAS for a Class of Stochastic Dynamic Programs}.
\newblock In {\em Automata, Languages, and Programming - 39th International
  Colloquium, {ICALP} 2018}, 2018.

\bibitem[HKL15]{HKL15}
Lisa Hellerstein, Devorah Kletenik, and Patrick Lin.
\newblock Discrete stochastic submodular maximization: Adaptive vs.
  non-adaptive vs. offline.
\newblock In {\em Algorithms and Complexity - 9th International Conference,
  {CIAC} 2015, Paris, France, May 20-22, 2015. Proceedings}, pages 235--248,
  2015.

\bibitem[LPRY08]{LiuPRY-SIGMOD08}
Zhen Liu, Srinivasan Parthasarathy, Anand Ranganathan, and Hao Yang.
\newblock Near-optimal algorithms for shared filter evaluation in data stream
  systems.
\newblock In {\em Proceedings of the {ACM} {SIGMOD} International Conference on
  Management of Data, {SIGMOD} 2008, Vancouver, BC, Canada, June 10-12, 2008},
  pages 133--146, 2008.

\bibitem[LY13]{LiYuan-STOC13}
Jian Li and Wen Yuan.
\newblock Stochastic combinatorial optimization via poisson approximation.
\newblock In {\em Symposium on Theory of Computing Conference, STOC'13, Palo
  Alto, CA, USA, June 1-4, 2013}, pages 971--980, 2013.

\bibitem[Ma14]{Ma-SODA14}
Will Ma.
\newblock Improvements and generalizations of stochastic knapsack and
  multi-armed bandit approximation algorithms: Extended abstract.
\newblock In {\em SODA}, pages 1154--1163, 2014.

\bibitem[Mes06]{mestre2006greedy}
Juli{\'a}n Mestre.
\newblock Greedy in approximation algorithms.
\newblock In {\em European Symposium on Algorithms}, pages 528--539. Springer,
  2006.

\bibitem[RS17]{RS-SODA17}
Aviad Rubinstein and Sahil Singla.
\newblock Combinatorial prophet inequalities.
\newblock In {\em Proceedings of the Twenty-Eighth Annual ACM-SIAM Symposium on
  Discrete Algorithms}, pages 1671--1687. SIAM, 2017.

\bibitem[RS{\"U}05]{RSU-JET05}
Alvin~E.\ Roth, Tayfun S{\"o}nmez, and M.\~Utku {\"U}nver.
\newblock Pairwise kidney exchange.
\newblock {\em J. Econom. Theory}, 125(2):151--188, 2005.

\end{thebibliography}

\end{document}